# Anomalous Maxwell-Garnett theory for photonic time crystals

Zheng Gong,[1,2] Ruoxi Chen,[1,2] Hongsheng Chen,[1,2,*] and Xiao Lin[1,2,*]

[1]*Interdisciplinary Center for Quantum Information, State Key Laboratory of Extreme Photonics and Instrumentation, Zhejiang Key Laboratory of Intelligent Electromagnetic Control and Advanced Electronic Integration, College of Information Science & Electronic Engineering, Zhejiang University, Hangzhou 310027, China.*

[2]*International Joint Innovation Center, The Electromagnetics Academy at Zhejiang University, Zhejiang University, Haining 314400, China.*

[*]*Corresponding authors. Email: xiaolinzju@zju.edu.cn (X. Lin); hansomchen@zju.edu.cn (H. Chen).*

**Maxwell-Garnett theory, dating back to James Clerk Maxwell-Garnett's foundational work in 1904, provides a simple yet powerful framework to describe the inhomogeneous structure as an effective homogeneous medium, which significantly reduces the overall complexity of analysis, calculation, and design. As such, the Maxwell-Garnett theory enables many practical applications in diverse realms, ranging from photonics, acoustics, mechanics, thermodynamics, to material science. It has long been thought that the Maxwell-Garnett theory of light in impedance-mismatched periodic structures is valid only *within* the long-wavelength limit, necessitating either the temporal or spatial period of light to be much larger than that of structures. Here, we break this long-held belief by revealing an anomalous Maxwell-Garnett theory for impedance-mismatched photonic time crystals *beyond* this long-wavelength limit. The key to this anomaly lies in the Fabry-Pérot resonance. We discover that under the Fabry-Pérot resonance, the impedance-mismatched photonic time crystal could be essentially equivalent to a homogeneous temporal slab simultaneously at specific discrete wavelengths, despite the temporal period of these light being comparable to or even much smaller than that of photonic time crystals.**

## Introduction

Maxwell-Garnett theory, as the simplest effective medium theory, is well-known for its enticing capability to model the inhomogeneous structure, such as metamaterials and random media with irregular geometries, as an effective homogeneous medium [1,2]. It was firstly proposed for light by James Clerk Maxwell-Garnett one century ago [3,4] and later generalized to other wave systems [5-9], including acoustic waves and water waves. According to the Maxwell-Garnett theory of light, the optical response of effective media could be formulated in a local or wavevector-independent fashion, which is solely governed by the geometrical parameters (e.g. filling ratio) and the intrinsic properties (e.g. impedance) of each constituent material. This mathematical simplification and physical elegance significantly reduce the computational cost and complexity, which would otherwise be computationally prohibitive, and further make the Maxwell-Garnett theory feasible to perform accurate analysis and design for intricate inhomogeneous structures with desired properties that are hard or even impossible to find in nature [10-13]. Therefore, the Maxwell-Garnett theory of light could greatly facilitate the flexible manipulation of light-matter interactions and is of fundamental importance to many practical applications, ranging from hyperlenses, metalenses, invisibility cloak, superscatterers, to absorbers [14-18].

Despite the long research history of effective medium theory [19-26], it is widely believed that the local Maxwell-Garnett theory of light would break down for impedance-mismatched periodic structures beyond the long-wavelength limit. The underlying reason is that when the spatial or temporal period of light is comparable to or much smaller than that of structures, the effective medium theory generally needs to be re-formulated into a nonlocal or wavevector-dependent fashion, in order to incorporate the influence of high-order scattering of light [27-31]. The resultant nonlocal effective medium theory is much more complicated and accurate than the local Maxwell-Garnett theory, but it oftentimes loses the practical convenience for the straightforward analysis and design of complex structures. In 1988, Ref. [25] found that the local Maxwell-Garnett theory can perform well beyond the long-wavelength limit for spatially-inhomogeneous structures, via impedance matching at the Brewster angle. Recently, this finding was generalized to impedance-matched temporally-inhomogeneous photonic time crystals [26].

Here, we reveal a universal mechanism to enable the anomalous local Maxwell-Garnett theory for impedance-mismatched photonic time crystals beyond the long-wavelength limit, which breaks the above century-old belief. This anomalous Maxwell-Garnett theory is essentially attributed to the Fabry-Pérot resonance. Since the Fabry-Pérot resonance condition can be readily achieved through the structural design, without any fundamental structural limitation (e.g. temporal period of photonic time crystals [32-41]) and any fundamental material limitation (e.g. impedances of constituent materials), our revealed mechanism for the anomalous Maxwell-Garnett theory circumvents the critical requirement for photonic time crystals being either within the long-wavelength limit or impedance-matched. Therefore, our finding further develops the conventional Maxwell-Garnett theory and might be crucial to the continuous exploration of temporal or spatiotemporal media [42-51].

**Results**

We begin with the introduction of the Maxwell-Garnett theory for photonic time crystals in Fig. 1. Without loss of generality, the photonic time crystal is homogeneous in space but has a time period $T_{\mathrm{PTC}}$ and a temporal interface number $N$, where $N=\infty$ without specific specification, and it is composed of two constituent media in Fig. 1(a). The constituent medium X (X = I or II) has the temporal filling ratio $\tau_{\mathrm{X}}/T_{\mathrm{PTC}}$, the permittivity $\varepsilon_{\mathrm{X}}$, the permeability $\mu_{\mathrm{X}}$, and the impedance $\eta_{\mathrm{X}}=\sqrt{\mu_{\mathrm{X}}/\varepsilon_{\mathrm{X}}}$, where $T_{\mathrm{PTC}}=\tau_{\mathrm{I}}+\tau_{\mathrm{II}}$. According to the Bloch band theory, the dispersion relation of photonic time crystals can be analytically obtained as [52]

$$\cos(\omega_{\mathrm{PTC}}\cdot T_{\mathrm{PTC}})=\cos(\omega_{\mathrm{I}}\tau_{\mathrm{I}})\cos(\omega_{\mathrm{II}}\tau_{\mathrm{II}})-\frac{1}{2}\left(\frac{\eta_{\mathrm{I}}}{\eta_{\mathrm{II}}}+\frac{\eta_{\mathrm{II}}}{\eta_{\mathrm{I}}}\right)\sin(\omega_{\mathrm{I}}\tau_{\mathrm{I}})\sin(\omega_{\mathrm{II}}\tau_{\mathrm{II}}) \tag{1}$$

where $\omega_{\mathrm{PTC}}$ is the eigenfrequency of light inside the photonic time crystal, the wavevector $k=|\bar{k}|>0$ is a conservable quantity due to the momentum conservation in temporal media [53], and $\omega_{\mathrm{X}}=k/\sqrt{\mu_{\mathrm{X}}\varepsilon_{\mathrm{X}}}$ is the angular frequency of light in medium X.

When the local Maxwell-Garnett theory works, the designed photonic time crystal could in principle be effectively modelled as a homogeneous temporal medium with the permittivity $\varepsilon_{\mathrm{MG}}$ and the permeability $\mu_{\mathrm{MG}}$ in Fig. 1(b). Correspondingly, for the incident light with a given wavevector $k$, the eigenfrequency $\omega_{\mathrm{MG}}=2\pi/T_{\mathrm{MG}}=k/\sqrt{\mu_{\mathrm{MG}}\varepsilon_{\mathrm{MG}}}$ of light predicted by the Maxwell-Garnett theory should be equal to the eigenfrequency $\omega_{\mathrm{PTC}}$ calculated by the Bloch band theory, namely $\omega_{\mathrm{MG}}=\omega_{\mathrm{PTC}}$, where $T_{\mathrm{MG}}$ essentially corresponds to the temporal period of incident light. By substituting $\omega_{\mathrm{MG}}=\omega_{\mathrm{PTC}}$ into equation (1), we further have

$$\cos(\omega_{\mathrm{MG}}\cdot T_{\mathrm{PTC}})=\cos(2\pi\cdot T_{\mathrm{PTC}}/T_{\mathrm{MG}})=\cos(\omega_{\mathrm{I}}\tau_{\mathrm{I}})\cos(\omega_{\mathrm{II}}\tau_{\mathrm{II}})-\frac{1}{2}\left(\frac{\eta_{\mathrm{I}}}{\eta_{\mathrm{II}}}+\frac{\eta_{\mathrm{II}}}{\eta_{\mathrm{I}}}\right)\sin(\omega_{\mathrm{I}}\tau_{\mathrm{I}})\sin(\omega_{\mathrm{II}}\tau_{\mathrm{II}}) \tag{2}$$

Upon close inspection of equation (2), it might be simplified under three distinct conditions, which directly leads to the emergence to three distinct types of local Maxwell-Garnett theories.

For type 1, when $\omega_{\mathrm{MG}}T_{\mathrm{PTC}} = 2\pi \cdot T_{\mathrm{PTC}}/T_{\mathrm{MG}} \to 0$ and $\omega_{\mathrm{X}}\tau_{\mathrm{X}} \to 0$, the incident light is within the long-wavelength limit. When within this long-wavelength limit, the Taylor expansion is applicable to equation (2), namely $\cos(\omega_{\mathrm{MG}}T_{\mathrm{PTC}}) \approx 1-(\omega_{\mathrm{MG}}T_{\mathrm{PTC}})^2/2$, $\cos(\omega_{\mathrm{X}}\tau_{\mathrm{X}}) \approx 1-(\omega_{\mathrm{X}}\tau_{\mathrm{X}})^2/2$, and $\sin(\omega_{\mathrm{X}}\tau_{\mathrm{X}}) \approx \omega_{\mathrm{X}}\tau_{\mathrm{X}}$. This way, after some calculations, equation (2) can be reduced to

$$\frac{T_{\mathrm{PTC}}}{\mu_{\mathrm{MG}}} \cdot \frac{T_{\mathrm{PTC}}}{\varepsilon_{\mathrm{MG}}} = \left(\frac{\tau_{\mathrm{I}}}{\mu_{\mathrm{I}}} + \frac{\tau_{\mathrm{II}}}{\mu_{\mathrm{II}}}\right) \cdot \left(\frac{\tau_{\mathrm{I}}}{\varepsilon_{\mathrm{I}}} + \frac{\tau_{\mathrm{II}}}{\varepsilon_{\mathrm{II}}}\right) \tag{3}$$

Accordingly, one possible solution to equation (3) is

$$\begin{matrix} \dfrac{T_{\mathrm{PTC}}}{\varepsilon_{\mathrm{MG}}} = \dfrac{\tau_{\mathrm{I}}}{\varepsilon_{\mathrm{I}}} + \dfrac{\tau_{\mathrm{II}}}{\varepsilon_{\mathrm{II}}} \\ \dfrac{T_{\mathrm{PTC}}}{\mu_{\mathrm{MG}}} = \dfrac{\tau_{\mathrm{I}}}{\mu_{\mathrm{I}}} + \dfrac{\tau_{\mathrm{II}}}{\mu_{\mathrm{II}}} \end{matrix}, \text{ if \textit{within} the long-wavelength limit (including } \omega_{\mathrm{MG}}T_{\mathrm{PTC}} \to 0) \tag{4}$$

Equation (4) is exactly the conventional Maxwell-Garnett mixing formulas [1,2], widely known for temporal media [21]. Generally, this conventional Maxwell-Garnett theory governed by equation (4) can obtain $\omega_{\mathrm{MG}} = \omega_{\mathrm{PTC}}$ (or more precisely speaking, $\omega_{\mathrm{MG}} \approx \omega_{\mathrm{PTC}}$) only within the long-wavelength limit, as shown in Figs. 2(a) and 2(b).

For type 2, when $\eta_{\mathrm{I}} = \eta_{\mathrm{II}}$, the designed photonic time crystal is impedance-matched. Accordingly, the impedance $\eta_{\mathrm{MG}} = \sqrt{\mu_{\mathrm{MG}}/\varepsilon_{\mathrm{MG}}}$ of effective temporal medium is the same as that of each constituent material, namely $\eta_{\mathrm{MG}} = \eta_{\mathrm{I}} = \eta_{\mathrm{II}}$. By substituting this impedance-matching condition into equation (2), equation (2) can be simplified to $\cos(\omega_{\mathrm{MG}}T_{\mathrm{PTC}}) = \cos(\omega_{\mathrm{I}}\tau_{\mathrm{I}})\cos(\omega_{\mathrm{II}}\tau_{\mathrm{II}}) - \sin(\omega_{\mathrm{I}}\tau_{\mathrm{I}})\sin(\omega_{\mathrm{II}}\tau_{\mathrm{II}}) = \cos(\omega_{\mathrm{I}}\tau_{\mathrm{I}} + \omega_{\mathrm{II}}\tau_{\mathrm{II}})$. This way, one simple solution to equation (2) is $\omega_{\mathrm{MG}}T_{\mathrm{PTC}} = \omega_{\mathrm{I}}\tau_{\mathrm{I}} + \omega_{\mathrm{II}}\tau_{\mathrm{II}}$, namely

$$\frac{T_{\mathrm{PTC}}}{\sqrt{\mu_{\mathrm{MG}}\varepsilon_{\mathrm{MG}}}} = \frac{\tau_{\mathrm{I}}}{\sqrt{\mu_{\mathrm{I}}\varepsilon_{\mathrm{I}}}} + \frac{\tau_{\mathrm{II}}}{\sqrt{\mu_{\mathrm{II}}\varepsilon_{\mathrm{II}}}} \tag{5}$$

By combining the impedance-matching condition and equation (5), we further have

$$\begin{matrix} \dfrac{T_{\mathrm{PTC}}}{\varepsilon_{\mathrm{MG}}} = \dfrac{\tau_{\mathrm{I}}}{\varepsilon_{\mathrm{I}}} + \dfrac{\tau_{\mathrm{II}}}{\varepsilon_{\mathrm{II}}} \\ \dfrac{T_{\mathrm{PTC}}}{\mu_{\mathrm{MG}}} = \dfrac{\tau_{\mathrm{I}}}{\mu_{\mathrm{I}}} + \dfrac{\tau_{\mathrm{II}}}{\mu_{\mathrm{II}}} \end{matrix}, \text{ if } \eta_{\mathrm{I}} = \eta_{\mathrm{II}}, \text{ for } \forall\ \omega_{\mathrm{MG}}T_{\mathrm{PTC}}/2\pi = T_{\mathrm{PTC}}/T_{\mathrm{MG}} \tag{6}$$

The anomalous local Maxwell-Garnett theory [26] governed by equation (6) is in accordance with the conventional one governed by equation (4), but it can now perform well and obtain $\omega_{\mathrm{MG}} = \omega_{\mathrm{PTC}}$ exactly by exploiting the impedance matching despite beyond the long-wavelength limit, as shown in Figs. 2(a) and 2(c).

For type 3, when $\sin(\omega_{\mathrm{I}}\tau_{\mathrm{I}}) = 0$ or $\sin(\omega_{\mathrm{II}}\tau_{\mathrm{II}}) = 0$, one constituent medium (e.g. medium I used in the calculation below) of photonic time crystals has the temporal Fabry-Pérot resonance. Under the scenario of Fabry-Pérot resonance of medium I, the amplitude of light transmitting through medium I remains unchanged [54]. In other words, medium I would not contribute to the impedance of the effective temporal medium. Accordingly, the impedance $\eta_{\mathrm{MG}}$ of the effective temporal medium could be the same as the other constituent medium (i.e. medium II) of photonic time crystals, namely $\eta_{\mathrm{MG}} = \eta_{\mathrm{II}}$. By substituting these conditions of $\sin(\omega_{\mathrm{I}}\tau_{\mathrm{I}}) = 0$ (i.e. $\cos(\omega_{\mathrm{I}}\tau_{\mathrm{I}}) = (-1)^m$ and $\omega_{\mathrm{I}}\tau_{\mathrm{I}} = m\pi$, $m \in \mathbb{N}$) and $\eta_{\mathrm{MG}} = \eta_{\mathrm{II}}$ into equation (2), equation (2) can be reduced to $\cos(\omega_{\mathrm{MG}}T_{\mathrm{PTC}}) = \pm\cos(\omega_{\mathrm{II}}\tau_{\mathrm{II}}) = \cos(m\pi + \omega_{\mathrm{II}}\tau_{\mathrm{II}}) = \cos(\omega_{\mathrm{I}}\tau_{\mathrm{I}} + \omega_{\mathrm{II}}\tau_{\mathrm{II}})$. This way, one possible solution to equation (2) is $\omega_{\mathrm{MG}}T_{\mathrm{PTC}} = \omega_{\mathrm{I}}\tau_{\mathrm{I}} + \omega_{\mathrm{II}}\tau_{\mathrm{II}}$, namely

$$\frac{T_{\mathrm{PTC}}}{\sqrt{\mu_{\mathrm{MG}}\varepsilon_{\mathrm{MG}}}} = \frac{\tau_{\mathrm{I}}}{\sqrt{\mu_{\mathrm{I}}\varepsilon_{\mathrm{I}}}} + \frac{\tau_{\mathrm{II}}}{\sqrt{\mu_{\mathrm{II}}\varepsilon_{\mathrm{II}}}} \tag{7}$$

Since $\omega_{\mathrm{MG}}T_{\mathrm{PTC}} = \omega_{\mathrm{I}}\tau_{\mathrm{I}} + \omega_{\mathrm{II}}\tau_{\mathrm{II}} > \omega_{\mathrm{I}}\tau_{\mathrm{I}} = m\pi \geq \pi$, we directly have $\omega_{\mathrm{MG}}T_{\mathrm{PTC}}/2\pi = T_{\mathrm{PTC}}/T_{\mathrm{MG}} > 1/2$, indicating the temporal Fabry-Pérot resonance occurs only beyond the long-wavelength limit.

By further combining $\eta_{\mathrm{MG}} = \eta_{\mathrm{II}}$ and equation (7), a slightly modified but still local-form Maxwell-Garnett mixing formulas can be obtained as follows

$$\begin{aligned} \frac{T_{\mathrm{PTC}}}{\varepsilon_{\mathrm{MG}}} &= \frac{\tau_{\mathrm{I}}}{\varepsilon_{\mathrm{I}}\eta_{\mathrm{I}}/\eta_{\mathrm{II}}} + \frac{\tau_{\mathrm{II}}}{\varepsilon_{\mathrm{II}}} \\ \frac{T_{\mathrm{PTC}}}{\mu_{\mathrm{MG}}} &= \frac{\tau_{\mathrm{I}}}{\mu_{\mathrm{I}}\eta_{\mathrm{II}}/\eta_{\mathrm{I}}} + \frac{\tau_{\mathrm{II}}}{\mu_{\mathrm{II}}} \end{aligned}, \text{ if } \sin(\omega_{\mathrm{I}}\tau_{\mathrm{I}}) = 0, \text{ for } \omega_{\mathrm{MG}}T_{\mathrm{PTC}}/2\pi = T_{\mathrm{PTC}}/T_{\mathrm{MG}} > 1/2 \tag{8}$$

Remarkably, this anomalous Maxwell-Garnett theory via the Fabry-Pérot resonance can obtain $\omega_{\mathrm{MG}} = \omega_{\mathrm{PTC}}$ exactly at specific discrete frequencies (i.e. $\omega_{\mathrm{MG}} = m\pi\sqrt{\mu_{\mathrm{I}}\varepsilon_{\mathrm{I}}}/(\tau_{\mathrm{I}}\sqrt{\mu_{\mathrm{MG}}\varepsilon_{\mathrm{MG}}})$) beyond the long-wavelength limit, as shown in Figs. 2(a) and 2(d), but without resorting to the impedance-matching condition. On the other hand, the anomalous Maxwell-Garnett theory via the impedance matching always requires the existence of magnetic response, namely either $\mu_{\mathrm{I}} \neq \mu_0$ or $\mu_{\mathrm{II}} \neq \mu_0$, and it is thus applicable to only magnetic photonic time crystals with $\mu_{\mathrm{I}} \neq \mu_{\mathrm{II}}$. By contrast, our revealed anomalous Maxwell-Garnett theory via the Fabry-Pérot resonance does not have any fundamental material constraint and is applicable to both magnetic and non-magnetic (i.e. $\mu_{\mathrm{I}} = \mu_{\mathrm{II}} = \mu_0$) photonic time crystals. We highlight that our revealed anomalous Maxwell-Garnett theory via the Fabry-Pérot resonance has never been discussed before.

In addition to check the criterion of $\omega_{\mathrm{MG}} = \omega_{\mathrm{PTC}}$, another criterion to examine the accuracy of Maxwell-Garnett theory is to check the equivalence between the transmission coefficient $\tilde{t}_{\mathrm{PTC}}$ (or the reflection coefficient $\tilde{r}_{\mathrm{PTC}}$) for a temporally finitely-thick photonic time crystal and that ($\tilde{t}_{\mathrm{MG}}$ or $\tilde{r}_{\mathrm{MG}}$) for the effective temporal slab, namely $\tilde{t}_{\mathrm{PTC}} = \tilde{t}_{\mathrm{MG}}$ (or $\tilde{r}_{\mathrm{PTC}} = \tilde{r}_{\mathrm{MG}}$). By following this thought, we show the relative error $||\tilde{t}_{\mathrm{MG}}|^2 - |\tilde{t}_{\mathrm{PTC}}|^2|/|\tilde{t}_{\mathrm{PTC}}|^2$of the energy transmittivity in the $\eta_{\mathrm{I}}/\eta_{\mathrm{II}}$ - $kcT_{\mathrm{PTC}}/2\pi$ parameter space in Figs. 3(a) and 3(b). For illustration, these designed photonic time crystals are surrounded by temporally semi-infinite vacuum. For magnetic photonic time crystals in Fig. 3(a), we have $||\tilde{t}_{\mathrm{MG}}|^2 - |\tilde{t}_{\mathrm{PTC}}|^2|/|\tilde{t}_{\mathrm{PTC}}|^2 \to 0$ and then $\tilde{t}_{\mathrm{PTC}} \approx \tilde{t}_{\mathrm{MG}}$ in the regime with $kcT_{\mathrm{PTC}}/2\pi = \omega_{\mathrm{MG}}T_{\mathrm{PTC}}/2\pi \cdot \sqrt{(\mu_{\mathrm{MG}}/\mu_0)(\varepsilon_{\mathrm{MG}}/\varepsilon_0)}$. The first scenario indicates that the conventional Maxwell-Garnett theory remains valid only within the long-wavelength limit, as schematically shown in Fig. 3(c). Meanwhile, we have $||\tilde{t}_{\mathrm{MG}}|^2 - |\tilde{t}_{\mathrm{PTC}}|^2|/|\tilde{t}_{\mathrm{PTC}}|^2 = 0$ and $\tilde{t}_{\mathrm{PTC}} = \tilde{t}_{\mathrm{MG}}$ in Fig. 3(a) in the regime with $\eta_{\mathrm{I}} = \eta_{\mathrm{II}}$ in Fig. 3(a), for arbitrary frequencies of incident light. The second scenario verifies the accuracy of the anomalous Maxwell-Garnett theory via the impedance matching [26]. For non-magnetic photonic time crystals in Fig. 3(b), we have $\tilde{t}_{\mathrm{PTC}} = \tilde{t}_{\mathrm{MG}}$ at a series of Fabry-Pérot resonant lines governed by $\sin(k\tau_{\mathrm{I}}/\sqrt{\mu_{\mathrm{I}}\varepsilon_{\mathrm{I}}}) = 0$ in the investigated parameter space. The third scenario essentially shows the existence of our revealed anomalous Maxwell-Garnett theory via the temporal Fabry-Pérot resonance. Remarkably, both types of anomalous Maxwell-Garnett theories could remain valid beyond the long-wavelength limit, as schematically illustrated in Fig. 3(d).

To facilitate further understanding, we show in Figs. 4(a) ,4(c) and 4(e) the spatiotemporal evolution of space-time wave packets interacting with various photonic time crystals beyond the long-wavelength limit. For conceptual brevity, these designed photonic time crystals are now surrounded by temporally semi-infinite media with the permittivity $\varepsilon_{\mathrm{MG}}$ and the permeability $\mu_{\mathrm{MG}}$. Moreover, for the direct comparison, we also show the spatiotemporal evolution of space-time wave packets interacting with the homogenized temporal slab of each photonic time crystal in Figs. 4(b), 4(d), and 4(f), respectively. In addition, since the anomalous Maxwell-Garnett theory via the impedance matching could remain valid for arbitrary frequency of light, the incident space-time wave packet is set to follow a continuous Gaussian-type waveform in Figs. 4(c) and 4(d). Similarly, the incident space-time wave packet is set to follow a multiple spatial harmonic waveform or a single spatial harmonic waveform in Figs. 4(e) and 4(f), since the anomalous Maxwell-Garnett theory via the temporal Fabry-Pérot resonance remains valid at specific discrete Fabry-Pérot resonant frequencies of light.

Under these judicious design in Fig. 4, there are at least two rules to follow, if the corresponding Maxwell-Garnett theory is valid. One rule is that there should be no reflection or no backward propagating light in the temporal region (i.e. the surrounding environment) behind photonic time crystals. The other rule to follow is that the spatiotemporal evolution of light in the temporal region behind the realistic photonic time crystal should be the same as that behind the homogenized temporal slab. According to these two rules of thumb, the conventional Maxwell-Garnett theory for conventional impedance-mismatched photonic time crystals in Figs. 4(a) and 4(b) generally breaks down beyond the long-wavelength limit. By contrast, the anomalous Maxwell-theory for either the impedance-matched photonic time crystal in Figs. 4(c) and 4(d) or the impedance-mismatched photonic time crystal with the temporal Fabry-Pérot resonance in Figs. 4(e) and 4(f) remains valid beyond the long-wavelength limit. On the other hand, we note that the backward-propagating waves could emerge inside the photonic time crystal, but they would further undergo the complete destructive interference when passing through the photonic time crystal, as exemplified by Figs. 4(e) and 4(f). The deviation between the spatiotemporal evolution of light inside the realistic photonic time crystal and that inside the homogenized temporal slab should not affect the validity of Maxwell-Garnett theory.

**Discussion**

In conclusion, we have found the existence of the anomalous Maxwell-Garnett theory for impedance-mismatched photonic time crystals beyond the long-wavelength limit by leveraging the temporal Fabry-Pérot resonance. Perhaps even more crucial is the vision emphasized by our finding: that this anomalous Maxwell-Garnett theory of light might be extended to spatially-inhomogeneous photonic crystals via the spatial Fabry-Pérot resonance, and that the analogous Maxwell-Garnett theory might exist in other wave systems, such as acoustic and water waves. Due to the mathematical simplicity and the physical elegance, our revealed anomalous Maxwell-Garnett theory of light may further stimulate the continuous exploration of more exotic light-matter interactions in temporal or spatiotemporal media [55-59], particularly in systems involving moving free electrons [60-70] or complex dipolar sources [71-75]. Moreover, our finding may intrigue the further exploration of many enticing open scientific questions that remain elusive, for example, the possible realization of broadband interfacial Cherenkov radiation from periodic structures. As background, the interfacial Cherenkov radiation [64], as originating from the interaction between free electrons and periodic structures, provides a disruptive way to create the directional light emission at arbitrary frequencies and is vital for the development of many enticing on-chip applications, such as integrated light sources at previously hard-to-reach frequencies and miniaturized particle detectors with enhanced sensitivity. However, the interfacial Cherenkov radiation severely suffers from the chromatic issue, due to the inherent structural dispersion of periodic structures. Whether it is possible to achieve the achromatic interfacial Cherenkov radiation from periodic structures via the anomalous Maxwell-Garnett theory of light is certainly worthy of in-depth exploration.

## Methods

**Accuracy of Maxwell-Garnett theory in predicting the transmission and reflection coefficients of photonic time crystal.** Under the condition of three types of Maxwell-Garnett theories, we rigorously prove in supplementary section S1 that the transmission coefficient $\tilde{t}_{\mathrm{PTC}}$ and the reflection coefficient $\tilde{r}_{\mathrm{PTC}}$ for a temporally finitely-thick photonic time crystal, are equivalent to those ($\tilde{t}_{\mathrm{MG}}$ and $\tilde{r}_{\mathrm{MG}}$) for the effective temporal slab.

**Spatiotemporal evolution of various wave packets interacting with photonic time crystals beyond the long-wavelength limit.** The spatiotemporal evolution of various wave packets, represented by the field distribution in the space-time ($z$ - $t$) diagram, is analytically obtained in supplementary section S2.

## Author contributions

Z.G. and X.L. conceived the idea; Z.G. performed the calculation; R.C., H.C., and X.L. helped to analyze the data; Z.G. and X.L. wrote the paper; H.C. and X.L. supervised the project.

## Conflict of interest

The authors have no conflicts to disclose.

## Data availability

All theoretical and numerical findings can be reproduced based on the information in the article and/or supplementary sections. The data represented in all figures are available on xxxx.

## Acknowledgement

X.L. acknowledges the support partly from the Fundamental Research Funds for the Central Universities under Grant No. 226-2024-00022, the National Natural Science Fund for Excellent Young Scientists Fund Program (Overseas) of China, the National Natural Science Foundation of China (NSFC) under Grant No. 62475227 and No. 62175212, and Zhejiang Provincial Natural Science Fund Key Project under Grant No. LZ23F050003. H.C. acknowledges the support from the Key Research and Development Program of the Ministry of Science and Technology under Grants No. 2022YFA1404704, 2022YFA1405200, and 2022YFA1404902, the Key Research and Development Program of Zhejiang Province under Grant No. 2022C01036, and the Fundamental Research Funds for the Central Universities. R.C. acknowledges the support from the NSFC under Grant No. 623B2089.

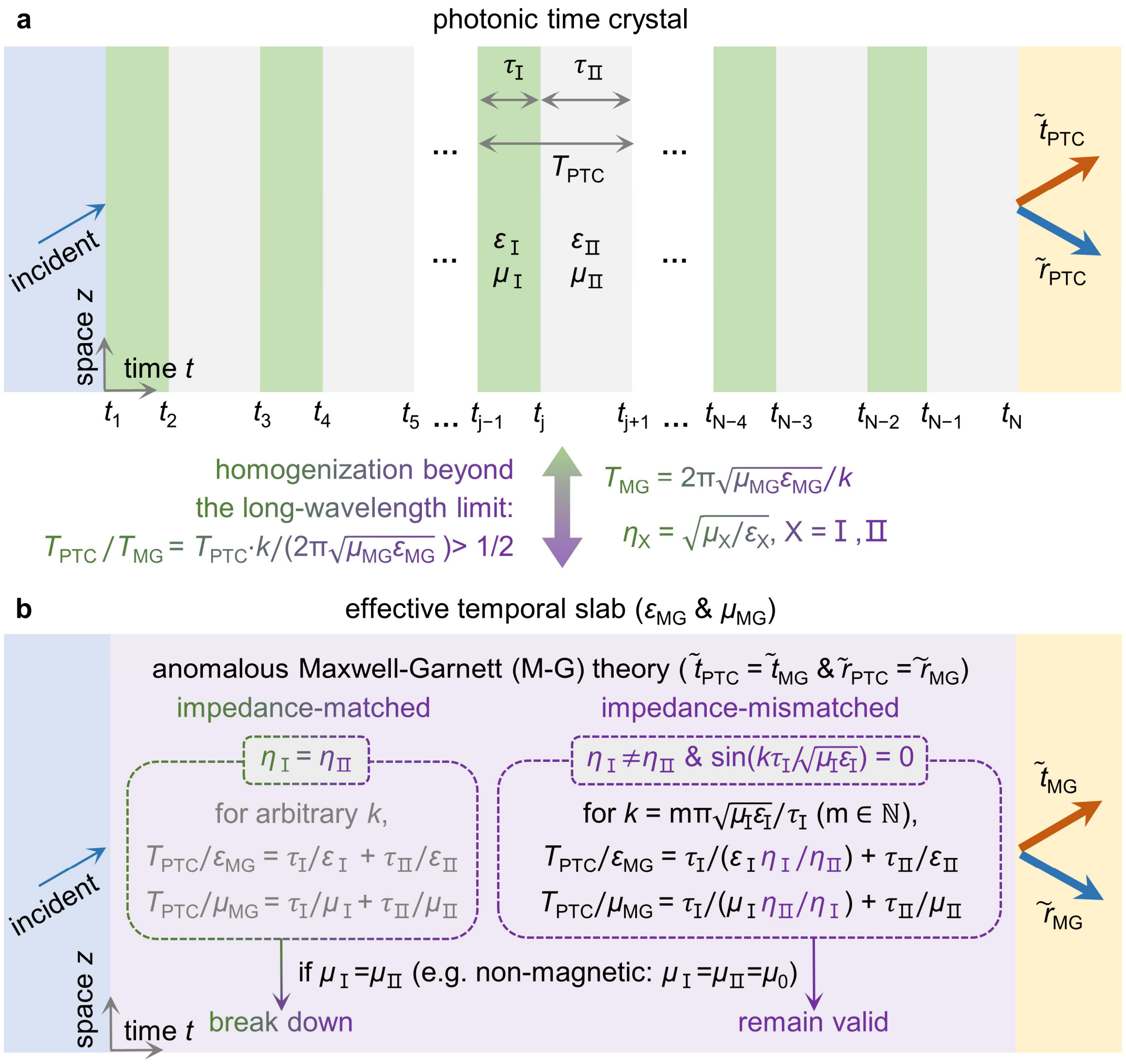

**FIG. 1.** Conceptual illustration of anomalous Maxwell-Garnett theory for impedance-mismatched photonic time crystals beyond the long-wavelength limit. (a) Structural schematic of a spatially homogeneous photonic time crystal with $N$ temporal interfaces. The $j$-th temporal interface is created by a step change in permittivity and/or permeability at time $t = t_j$. The alternating constituent medium X (X = I or II) has a time duration $\tau_{\mathrm{X}}$, the permittivity $\varepsilon_{\mathrm{X}}$, the permeability $\mu_{\mathrm{X}}$, and the wave impedance $\eta_{\mathrm{X}} = \sqrt{\mu_{\mathrm{X}}/\varepsilon_{\mathrm{X}}}$. (b) Structural schematic of the effective temporal slab homogenized via the anomalous Maxwell-Garnett theory. Beyond the long-wavelength limit, the temporal period $T_{\mathrm{MG}} = 2\pi\sqrt{\mu_{\mathrm{MG}}\varepsilon_{\mathrm{MG}}}/k$ of light predicted by Maxwell-Garnett theory is comparable to or even smaller than the temporal period $T_{\mathrm{PTC}}$ of photonic time crystals, e.g. $T_{\mathrm{PTC}}/T_{\mathrm{MG}} > 1/2$, where $\mu_{\mathrm{MG}}$ and $\varepsilon_{\mathrm{MG}}$ are the permeability and permittivity of the effective homogenized temporal slab, and $k$ is the spatial frequency of incident light. When the anomalous Maxwell-Garnett theory works, the transmission and reflection coefficients (i.e. $\tilde{t}_{\mathrm{PTC}}$ and $\tilde{r}_{\mathrm{PTC}}$) for the photonic time crystal are the same as those (i.e. $\tilde{t}_{\mathrm{MG}}$ and $\tilde{r}_{\mathrm{MG}}$) for the effective temporal slab, respectively, namely $\tilde{t}_{\mathrm{PTC}} = \tilde{t}_{\mathrm{MG}}$ and $\tilde{r}_{\mathrm{PTC}} = \tilde{r}_{\mathrm{MG}}$.

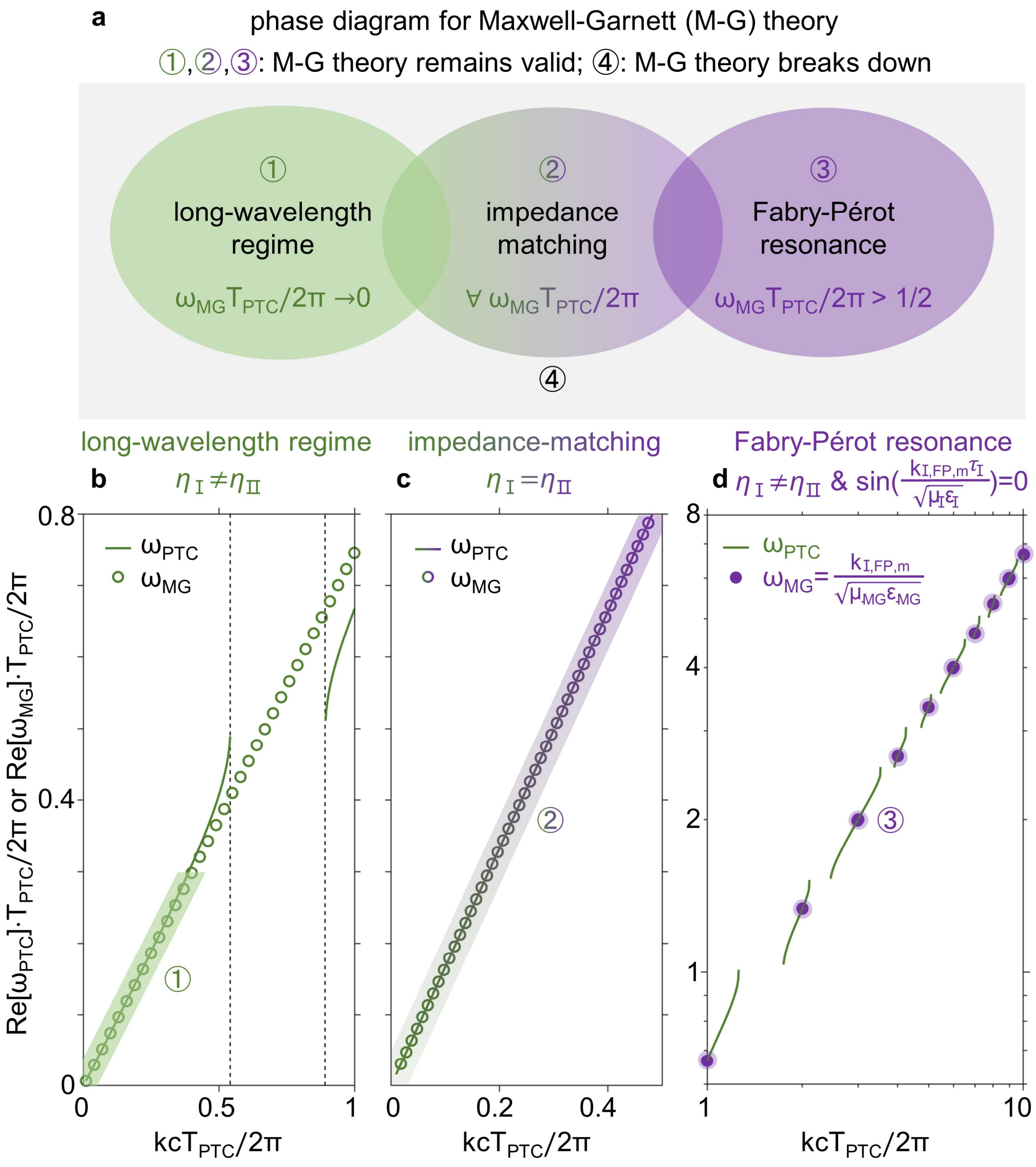

**FIG. 2.** Phase diagram for Maxwell-Garnett theory. (a) Classification of conventional and anomalous Maxwell-Garnett theories. While the conventional Maxwell-Garnett theory is limited to the long-wavelength regime ① (i.e. $T_{\mathrm{PTC}}/T_{\mathrm{MG}} = \omega_{\mathrm{MG}} T_{\mathrm{PTC}}/2\pi \to 0$), the anomalous Maxwell-Garnett theory remains valid beyond the long-wavelength limit by exploiting either the impedance matching (i.e. regime ② with $\forall\, \omega_{\mathrm{MG}} T_{\mathrm{PTC}}/2\pi$) or the Fabry-Pérot resonance (i.e. regime ③ with $\omega_{\mathrm{MG}} T_{\mathrm{PTC}}/2\pi > 1/2$), where $\omega_{\mathrm{MG}} = 2\pi/T_{\mathrm{MG}}$ is the eigenfrequency calculated via the Maxwell-Garnett theory. In regime ④, the Maxwell-Garnett theory breaks down. (b)-(d) Band structures of photonic time crystals judiciously designed to map various Maxwell-Garnett theories in (a). While the eigenfrequency $\omega_{\mathrm{PTC}}$ of photonic time crystals is multi-valued according to the Bloch theory, the branch cut of $\omega_{\mathrm{PTC}}$ closest to the frequency $\omega_{\mathrm{MG}}$ is chosen for comparison. For demonstration, $\varepsilon_{\mathrm{I}}/\varepsilon_0 = 1$, $\varepsilon_{\mathrm{II}}/\varepsilon_0 = 8.9$, and $\mu_{\mathrm{I}}/\mu_0 = \mu_{\mathrm{II}}/\mu_0 = 1$ are used in (b) and (d), while $\varepsilon_{\mathrm{I}}/\varepsilon_0 = 1$, $\varepsilon_{\mathrm{II}}/\varepsilon_0 = 8.9$, $\mu_{\mathrm{I}}/\mu_0 = 1/8.9$, and $\mu_{\mathrm{II}}/\mu_0 = 1$ are used in (c), where $\varepsilon_0$ and $\mu_0$ are the vacuum permittivity and permeability, and $c = 1/\sqrt{\mu_0\varepsilon_0}$ is the light speed in vacuum. Meanwhile, we set the wavelength in vacuum $\lambda_0 = cT_0 = 500$ nm, $\tau_{\mathrm{I}}/T_{\mathrm{PTC}} = \tau_{\mathrm{II}}/T_{\mathrm{PTC}} = 0.5$, the temporal period of photonic time crystals $T_{\mathrm{PTC}} = T_0$.

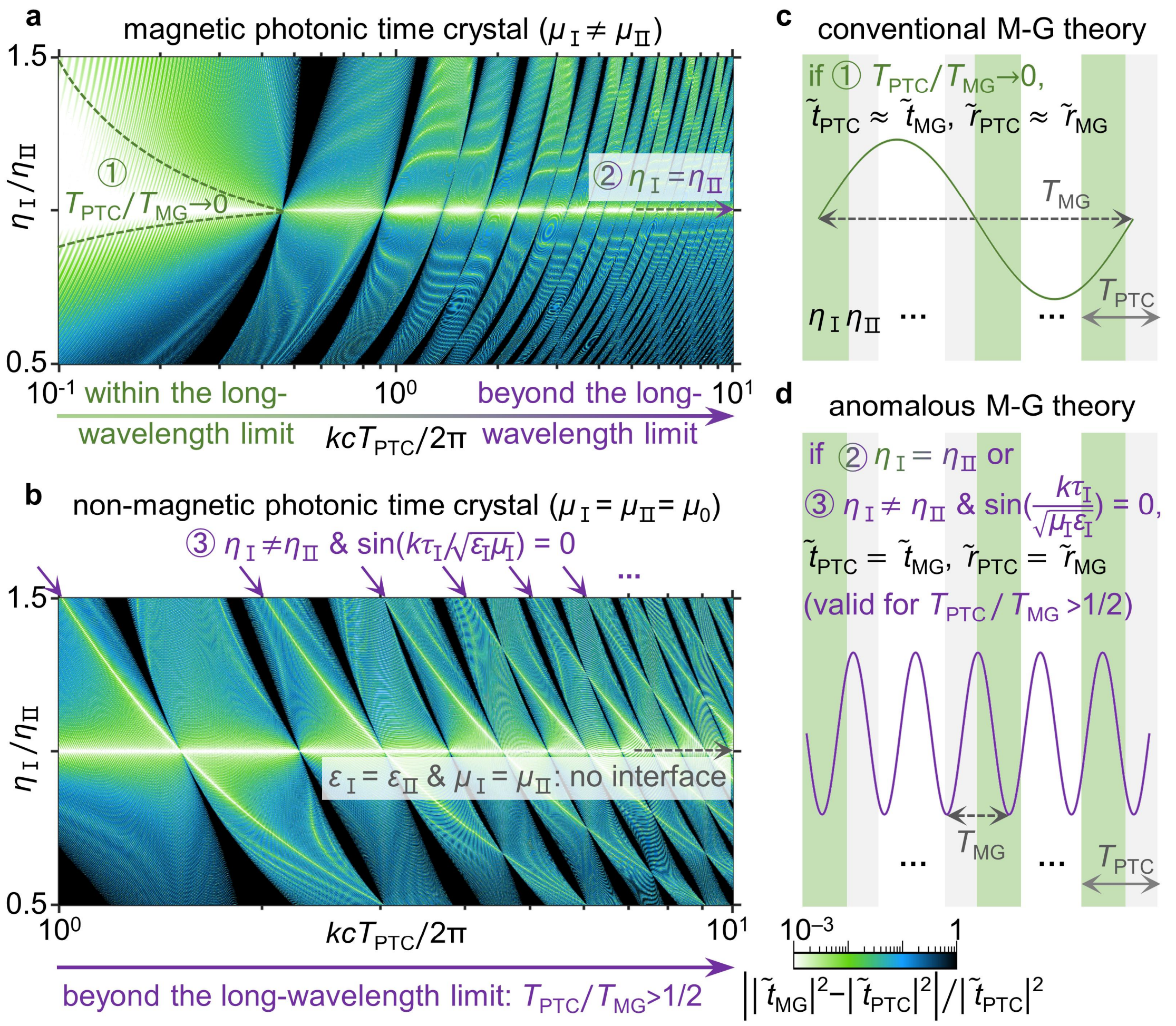

**FIG. 3.** Anomalous Maxwell-Garnett theory of light in the impedance-momentum parameter space. The photonic time crystal is surrounded by vacuum in the time domain and has a temporal interface number $N = 201$. (a),(b) $||\tilde{t}_{\mathrm{MG}}|^2 - |\tilde{t}_{\mathrm{PTC}}|^2|/|\tilde{t}_{\mathrm{PTC}}|^2$ as a function of $\eta_{\mathrm{I}}/\eta_{\mathrm{II}}$ and $kcT_{\mathrm{PTC}}/2\pi$. The relative error $||\tilde{t}_{\mathrm{MG}}|^2 - |\tilde{t}_{\mathrm{PTC}}|^2|/|\tilde{t}_{\mathrm{PTC}}|^2$ is used to quantitively describe the accuracy of Maxwell-Garnett theory in the homogenization of photonic time crystals. For illustration, $\varepsilon_{\mathrm{I}}/\varepsilon_0 = 1$, $\varepsilon_{\mathrm{II}}/\varepsilon_0 = 2.1$ and $\mu_{\mathrm{II}}/\mu_0 = 1$ are used in (a), while $\varepsilon_{\mathrm{II}}/\varepsilon_0 = 2.1$ and $\mu_{\mathrm{I}} = \mu_{\mathrm{II}} = \mu_0$ are used in (b). The temporal periods of photonic time crystals are the same as those in Fig. 2. For non-magnetic time crystals with $\eta_{\mathrm{I}}/\eta_{\mathrm{II}} = 1$ in (b), this trivial scenario directly corresponds to $\varepsilon_{\mathrm{I}} = \varepsilon_{\mathrm{II}}$ and $\mu_{\mathrm{I}} = \mu_{\mathrm{II}}$, indicating the absence of temporal interfaces inside the photonic time crystal. (c),(d) Comparison between conventional and anomalous Maxwell-Garnett theories. The conventional Maxwell-Garnett theory works only within the long-wavelength regime in (c), namely if $T_{\mathrm{PTC}}/T_{\mathrm{MG}} \to 0$. By contrast, the anomalous Maxwell-Garnett theory remains valid beyond the long-wavelength limit (e.g. $T_{\mathrm{PTC}}/T_{\mathrm{MG}} > 1/2$) in (d), either by exploiting the impedance matching or the Fabry-Pérot resonance.

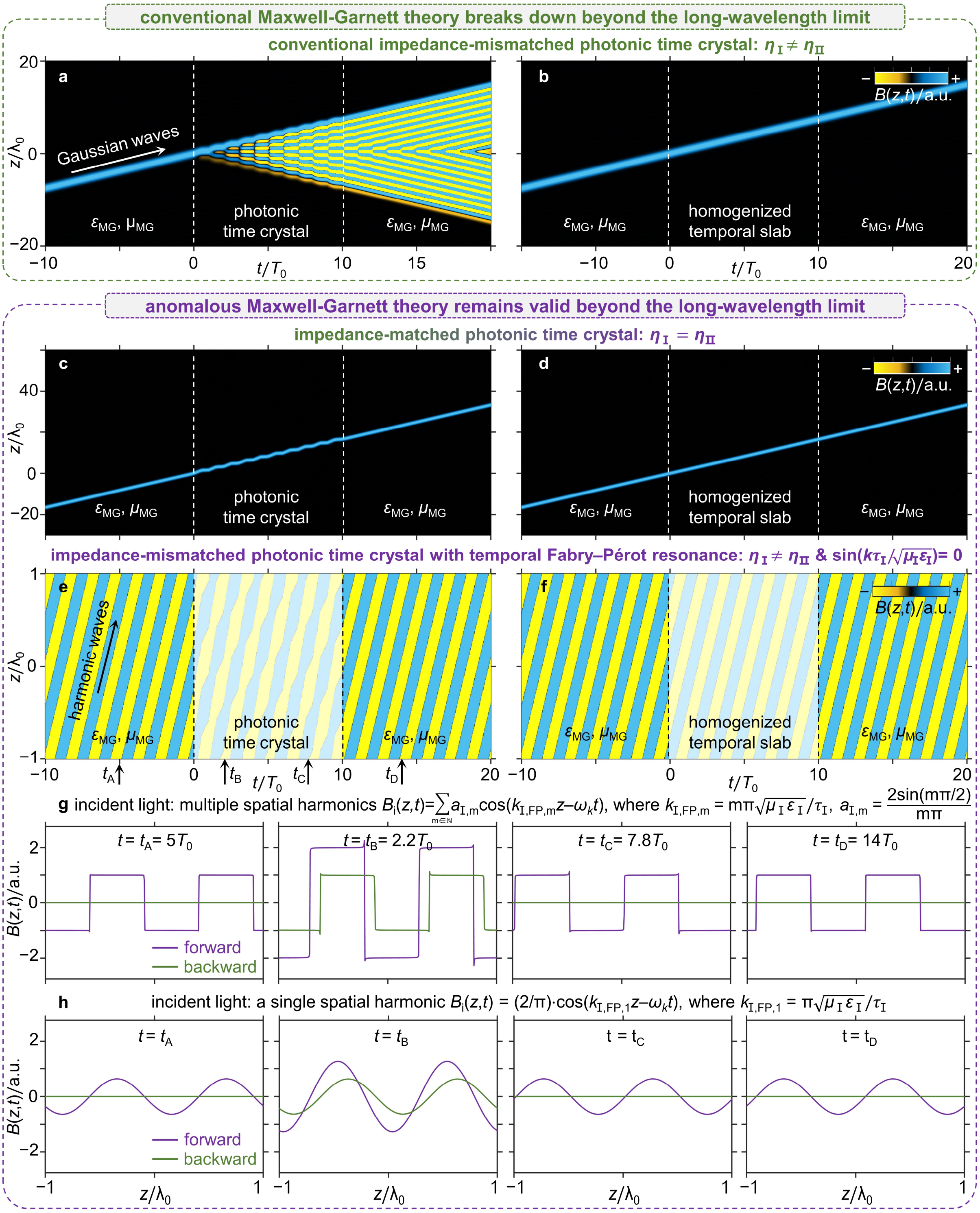

**FIG. 4.** Spatiotemporal evolution of space-time wave packets interacting with various photonic time crystals beyond the long-wavelength limit. For illustration, the photonic time crystal is surrounded by temporally semi-infinite media with the permittivity $\varepsilon_{MG}$ and the permeability $\mu_{MG}$. (a),(b) Conventional Maxwell-Garnett theory breaks down beyond the long-wavelength limit. (c)-(h) Anomalous Maxwell-Garnett theory remains valid beyond the long-wavelength limit. The photonic time crystal is impedance-mismatched (i.e.

$\eta_{\mathrm{I}} \neq \eta_{\mathrm{II}}$) in (a) and (e), but impedance-matched (i.e. $\eta_{\mathrm{I}} = \eta_{\mathrm{II}}$) in (c). Meanwhile, one constituent medium (e.g. medium I) in (e) satisfies the temporal Fabry- Pérot resonance condition, namely $\sin(\omega_{\mathrm{I}}\tau_{\mathrm{I}}) = 0$. The wave packet' states before entering, travelling inside, and after exiting the impedance-mismatched photonic time crystal with the temporal Fabry-Pérot resonance in (e) are highlighted in (g) and (h). The incident wave packet follows a Gaussian waveform $B_{\mathrm{i}}(z,t) = B(z,t < t_1) = \int_{-k_0}^{k_0} dk\, e^{-k^2/2\sigma_k^2} e^{ikz-i\omega t}$ in (a)-(d), a multiple-spatial-harmonic waveform $B_{\mathrm{i}}(z,t) = B(z,t < t_1) = \sum_{m\in\mathbb{N}} a_{\mathrm{I},m} \cos\left(k_{\mathrm{I,FP},m} z - \omega_k t\right)$ in (e)-(g), and a single-spatial-harmonic waveform $B(z,t < t_1) = \cos\left(k_{\mathrm{I,FP},1} r - \omega_k t\right)$ in (h), where $\sigma_k = k_0/3$, $k_0 = 2\pi/\lambda_0 = 2\pi/cT_0$, $k_{\mathrm{I,FP},m} = m\pi\sqrt{\mu_{\mathrm{I}}\varepsilon_{\mathrm{I}}}/\tau_{\mathrm{I}}$, $\omega_k = k/\sqrt{\mu\varepsilon}$, $a_{\mathrm{I},m} = \frac{2\sin(m\pi/2)}{m\pi}$, and $T_0 = T_{\mathrm{PTC}}$. The photonic time crystals in (a),(c),(e) are the same as those in Figs. 2(b)-2(d), respectively, except that we set the interface number $N = 21$ here.

Supplementary information for

# Anomalous Maxwell-Garnett theory for photonic time crystals

Zheng Gong, Ruoxi Chen, Hongsheng Chen, and Xiao Lin

## Guide To the Supplementary Sections

## S1 Accuracy of Maxwell-Garnett theory in predicting the transmission and reflection coefficients of photonic time crystal.

In this section, we analytically prove the accuracy of various Maxwell-Garnett theories, namely equations (4), (6), and (8) in the main text, in predicting the transmission and reflection coefficients of photonic time crystal; specifically, we show their equivalence with those in the effective temporal slab.

### S1.1 General formulation for space-harmonic fields

In this subsection, we start with the electromagnetic fields of a particular wavevector $k$ (e.g., electric displacement $\boldsymbol{D}_k$ and magnetic flux density $\boldsymbol{B}_k$) in the steady state, namely space-harmonic fields [1]. On this basis, Fourier theory can be applied to study the space-domain fields as follows

$$\begin{aligned} B(z,t) &= \int_{-\infty}^{+\infty} dk\ B_k(t)\cdot e^{ikz} \\ D(z,t) &= \int_{-\infty}^{+\infty} dk\ D_k(t)\cdot e^{ikz} \end{aligned} \tag{S1}$$

where we only consider a one-dimensional space $\boldsymbol{r}$ for conceptual brevity.

For the photonic time crystal with the structural setup in Fig. 1 in the main text, the permittivity and permeability in the whole space-time domain are given by

$$\varepsilon(t) = \varepsilon_j,\quad \mu(t) = \mu_j,\quad \text{region } j,\ 1 \le j \le N+1 \tag{S2}$$

where $N$ is the total temporal interface number; $\varepsilon_{\mathrm{MG}}$ and $\mu_{\mathrm{MG}}$ are the homogenized effective parameters. The field expressions for $\boldsymbol{D}_k$ and $\boldsymbol{B}_k$ (the subscript $k$ is neglected for concise expression) are assumed as follows

$$\begin{aligned} B(t) &= \begin{cases} a_1^+ e^{-i\omega_j(t-t_1)} & t \le t_1 \text{ (region 1)} \\ a_j^+ e^{-i\omega_j(t-t_{j-1})} + a_j^- e^{+i\omega_j(t-t_{j-1})} & t_{j-1} < t \le t_j \text{ (region } j) \\ a_{N+1}^+ e^{-i\omega_j(t-t_N)} + a_{N+1}^- e^{+i\omega_j(t-t_N)} & t_N < t \text{ (region } N+1) \end{cases} \\ D(t) &= \begin{cases} -\dfrac{1}{\eta_1} a_1^+ e^{-i\omega_j(t-t_1)} & t \le t_1 \\ -\dfrac{1}{\eta_j} a_j^+ e^{-i\omega_j(t-t_{j-1})} + \dfrac{1}{\eta_j} a_j^- e^{+i\omega_j(t-t_{j-1})} & t_{j-1} < t \le t_j \\ -\dfrac{1}{\eta_{N+1}} a_{N+1}^+ e^{-i\omega_j(t-t_N)} + \dfrac{1}{\eta_{N+1}} a_{N+1}^- e^{+i\omega_j(t-t_N)} & t_N < t \end{cases} \end{aligned} \tag{S3}$$

where $a_j^+$ ($a_j^-$) is the amplitude of the forward (backward) propagating wave components, and $\eta_j$ and $\omega_j$ are the wave impedance and wave frequency given by

$$\eta_j = \frac{\mu_j \omega_j}{k} = \sqrt{\mu_j/\varepsilon_j}\ ,\quad \omega_j = k\Big/\sqrt{\mu_j\varepsilon_j}\ ,\quad \forall j \tag{S4}$$

## S1.2 General formulation for the temporal characteristic matrix

In this subsection, we derive the characteristic matrix $\overline{\overline{M}}_j$ for a single temporal slab extending from $t = t_{j-1}$ to $t = t_j$, which relates the field values at its two temporal interfaces, namely

$$\begin{bmatrix} B_j(t_j) \\ D_j(t_j) \end{bmatrix} = \overline{\overline{M}}_j \begin{bmatrix} B_{j-1}(t_{j-1}) \\ D_{j-1}(t_{j-1}) \end{bmatrix} \tag{S5}$$

Equation (S5) is the generalization of Born's formulation for a single spatial slab [2] into the temporal case. The solution to $\overline{\overline{M}}_j$ can be obtained by enforcing temporal boundary condition and simple geometric optics. One the one hand, the continuity of the electric displacement $\boldsymbol{D}$ and magnetic flux density $\boldsymbol{B}$ before and after the temporal interface should be guaranteed, namely

$$\begin{bmatrix} B_{j-1}(t_{j-1}) \\ D_{j-1}(t_{j-1}) \end{bmatrix} = \begin{bmatrix} B_j(t_{j-1}) \\ D_j(t_{j-1}) \end{bmatrix}, \quad \forall j \in [2, N+1] \tag{S6}$$

Also note that the wavevector $k$ is a conservable quantity due to the boundary condition.

On the other hand, from the perspective of geometric optics by following equation (S3), one has

$$\begin{gathered} B_j(t_j) = e^{-i\omega_j \tau_j} a_j^+ + e^{+i\omega_j \tau_j} a_j^- \\ D_j(t_j) = -\frac{1}{\eta_j} e^{-i\omega_j \tau_j} a_j^+ + \frac{1}{\eta_j} e^{+i\omega_j \tau_j} a_j^- \\ \tau_j = t_j - t_{j-1} \end{gathered} \tag{S7}$$

where $\tau_j$ is the temporal duration of the slab in region $j$. The amplitudes $a_j^+$ and $a_j^-$ can be obtained by setting $t = t_{j-1}$ in equation (S3), and are related to $\boldsymbol{B}(t_j)$ and $\boldsymbol{D}(t_{j-1})$ by

$$\begin{bmatrix} a_j^+ \\ a_j^- \end{bmatrix} = \begin{bmatrix} 1/2 & -\eta_j/2 \\ 1/2 & \eta_j/2 \end{bmatrix} \begin{bmatrix} B_j(t_{j-1}) \\ D_j(t_{j-1}) \end{bmatrix} \tag{S8}$$

One can also write equation (S8) equivalently as

$$\begin{bmatrix} B_j(t_{j-1}) \\ D_j(t_{j-1}) \end{bmatrix} = \begin{bmatrix} 1 & 1 \\ -1/\eta_j & 1/\eta_j \end{bmatrix} \begin{bmatrix} a_j^+ \\ a_j^- \end{bmatrix} \tag{S9}$$

By substituting equation (S8) into equation (S7), and after some algebra, one has

$$\begin{bmatrix} B_j(t_j) \\ D_j(t_j) \end{bmatrix} = \begin{bmatrix} \cos(\omega_j \tau_j) & i\eta_j \sin(\omega_j \tau_j) \\ i\sin(\omega_j \tau_j)/\eta_j & \cos(\omega_j \tau_j) \end{bmatrix} \begin{bmatrix} B_j(t_{j-1}) \\ D_j(t_{j-1}) \end{bmatrix}, \quad \forall j \in [2, N+1] \tag{S10}$$

By combining formulas (S6) and (S10), one has the expression for temporal characteristic matrix as follows

$$\overline{\overline{M}}_j = \begin{bmatrix} \cos(\omega_j \tau_j) & i\eta_j \sin(\omega_j \tau_j) \\ i\sin(\omega_j \tau_j)/\eta_j & \cos(\omega_j \tau_j) \end{bmatrix} \tag{S11}$$

## S1.3 Temporal characteristic matrix for the photonic time crystal and the homogenized temporal slab

In this subsection, we obtain the temporal characteristic matrix $\overline{\overline{M}}_{\mathrm{PTC}}$ for the photonic time crystal and the homogenized temporal slab. For the photonic time crystal, we start with its unit-cell characteristic matrix as follows

$$\begin{aligned}\overline{\overline{M}}_{\mathrm{PTC,unit}} &= \begin{bmatrix}\overline{\overline{M}}_{\mathrm{PTC,unit,11}} & \overline{\overline{M}}_{\mathrm{PTC,unit,12}}\\ \overline{\overline{M}}_{\mathrm{PTC,unit,21}} & \overline{\overline{M}}_{\mathrm{PTC,unit,22}}\end{bmatrix}\\ &= \overline{\overline{M}}_{\mathrm{I}}\overline{\overline{M}}_{\mathrm{II}} = \begin{bmatrix}\cos(\omega_{\mathrm{I}}\tau_{\mathrm{I}}) & i\eta_{\mathrm{I}}\sin(\omega_{\mathrm{I}}\tau_{\mathrm{I}})\\ i\sin(\omega_{\mathrm{I}}\tau_{\mathrm{I}})/\eta_{\mathrm{I}} & \cos(\omega_{\mathrm{I}}\tau_{\mathrm{I}})\end{bmatrix}\begin{bmatrix}\cos(\omega_{\mathrm{II}}\tau_{\mathrm{II}}) & i\eta_{\mathrm{II}}\sin(\omega_{\mathrm{II}}\tau_{\mathrm{II}})\\ i\sin(\omega_{\mathrm{II}}\tau_{\mathrm{II}})/\eta_{\mathrm{II}} & \cos(\omega_{\mathrm{II}}\tau_{\mathrm{II}})\end{bmatrix}\end{aligned} \tag{S12}$$

After some algebra, one has

$$\begin{aligned}\overline{\overline{M}}_{\mathrm{PTC,unit,11}} &= \cos(\omega_{\mathrm{I}}\tau_{\mathrm{I}})\cos(\omega_{\mathrm{II}}\tau_{\mathrm{II}}) - \sin(\omega_{\mathrm{I}}\tau_{\mathrm{I}})\sin(\omega_{\mathrm{II}}\tau_{\mathrm{II}})\,\eta_{\mathrm{I}}/\eta_{\mathrm{II}}\\ \overline{\overline{M}}_{\mathrm{PTC,unit,12}} &= i\eta_{\mathrm{II}}\cos(\omega_{\mathrm{I}}\tau_{\mathrm{I}})\sin(\omega_{\mathrm{II}}\tau_{\mathrm{II}}) + i\eta_{\mathrm{I}}\sin(\omega_{\mathrm{I}}\tau_{\mathrm{I}})\cos(\omega_{\mathrm{II}}\tau_{\mathrm{II}})\\ \overline{\overline{M}}_{\mathrm{PTC,unit,21}} &= i/\eta_{\mathrm{I}}\sin(\omega_{\mathrm{I}}\tau_{\mathrm{I}})\cos(\omega_{\mathrm{II}}\tau_{\mathrm{II}}) + i/\eta_{\mathrm{II}}\cos(\omega_{\mathrm{I}}\tau_{\mathrm{I}})\sin(\omega_{\mathrm{II}}\tau_{\mathrm{II}})\\ \overline{\overline{M}}_{\mathrm{PTC,unit,22}} &= \cos(\omega_{\mathrm{I}}\tau_{\mathrm{I}})\cos(\omega_{\mathrm{II}}\tau_{\mathrm{II}}) - \sin(\omega_{\mathrm{I}}\tau_{\mathrm{I}})\sin(\omega_{\mathrm{II}}\tau_{\mathrm{II}})\,\eta_{\mathrm{II}}/\eta_{\mathrm{I}}\end{aligned} \tag{S13}$$

Note that $\overline{\overline{M}}_{\mathrm{unit}}$ is a unimodular matrix, then, with some knowledge from the matrix theory [2], one has

$$\begin{aligned}\overline{\overline{M}}_{\mathrm{PTC}} = \overline{\overline{M}}_{\mathrm{PTC,unit}}^{N_{\mathrm{unit}}} &= \begin{bmatrix}\overline{\overline{M}}_{\mathrm{PTC,unit,11}}U_{N_{\mathrm{unit}}-1}(a) - U_{N_{\mathrm{unit}}-2}(a) & \overline{\overline{M}}_{\mathrm{PTC,unit,12}}U_{N_{\mathrm{unit}}-1}(a)\\ \overline{\overline{M}}_{\mathrm{PTC,unit,21}}U_{N_{\mathrm{unit}}-1}(a) & \overline{\overline{M}}_{\mathrm{PTC,unit,22}}U_{N_{\mathrm{unit}}-1}(a) - U_{N_{\mathrm{unit}}-2}(a)\end{bmatrix}\\ a &= \frac{\overline{\overline{M}}_{\mathrm{PTC,unit,11}} + \overline{\overline{M}}_{\mathrm{PTC,unit,21}}}{2} = \cos(\omega_{\mathrm{I}}\tau_{\mathrm{I}} + \omega_{\mathrm{II}}\tau_{\mathrm{II}})\end{aligned} \tag{S14}$$

where $U_{N_{\mathrm{unit}}}(\cos x) = \sin\left[(N_{\mathrm{unit}} + 1)x\right]/\sin x$ represents the Chebyshev polynomials of the second kind.

For the homogenized temporal slab, it is easy to write its characteristic matrix as follows based on the derivation in last subsection.

$$\begin{aligned}\overline{\overline{M}}_{\mathrm{MG}} &= \begin{bmatrix}\cos(\omega_{\mathrm{MG}}\tau_{\mathrm{MG}}) & i\eta_{\mathrm{MG}}\sin(\omega_{\mathrm{MG}}\tau_{\mathrm{MG}})\\ i\sin(\omega_{\mathrm{MG}}\tau_{\mathrm{MG}})/\eta_{\mathrm{MG}} & \cos(\omega_{\mathrm{MG}}\tau_{\mathrm{MG}})\end{bmatrix}\\ \eta_{\mathrm{MG}} = \sqrt{\mu_{\mathrm{MG}}/\varepsilon_{\mathrm{MG}}}\ , &\quad \omega_{\mathrm{MG}} = k/\sqrt{\mu_{\mathrm{MG}}\varepsilon_{\mathrm{MG}}}\ , \quad \tau_{\mathrm{MG}} = (\tau_{\mathrm{I}} + \tau_{\mathrm{II}})\cdot N_{\mathrm{unit}}\end{aligned} \tag{S15}$$

where $\tau_{\mathrm{MG}}$, $\omega_{\mathrm{MG}}$, $\eta_{\mathrm{MG}}$ are the effective temporal duration, angular frequency and impedance of the temporal slab obtained via the Maxwell-Garnett mixing theory.

## S1.4 Transmission and reflection coefficients, and the energy transmittivity and reflectivity

In this subsection, we give a general derivation for the transmission and reflection coefficients $\tilde{t}$ and $\tilde{r}$, and the energy transmittivity and reflectivity for the photonic time crystal and the homogenized temporal slab.

For the photonic time crystal, $\tilde{t}_{\mathrm{PTC}}$ and $\tilde{r}_{\mathrm{PTC}}$ are defined as

$$\begin{aligned}\tilde{t}_{\mathrm{PTC}} &= a_{N+1}^{+}/a_1^{+} \\ \tilde{r}_{\mathrm{PTC}} &= a_{N+1}^{-}/a_1^{+}\end{aligned} \tag{S16}$$

Note here the transmission and reflection coefficients are defined with expect to the magnetic flux density $\boldsymbol{B}$. By the definition of the characteristic matrixes for equation (S5) and by combing equations (S8-S9), one has

$$\begin{bmatrix}1 & 1 \\ -1/\eta_{N+1} & 1/\eta_{N+1}\end{bmatrix}\begin{bmatrix}a_{N+1}^{+} \\ a_{N+1}^{-}\end{bmatrix} = \overline{\overline{M}}_{\mathrm{PTC}}\begin{bmatrix}1 & 1 \\ -1/\eta_1 & 1/\eta_1\end{bmatrix}\begin{bmatrix}a_1^{+} \\ a_1^{-}\end{bmatrix} \tag{S17}$$

After some calculation, one has the scattering matrix $\overline{\overline{S}}_{\mathrm{PTC}}$ for the photonic time crystal, namely

$$\begin{aligned}\begin{bmatrix}a_{N+1}^{+} \\ a_{N+1}^{-}\end{bmatrix} &= \overline{\overline{S}}_{\mathrm{PTC}}\begin{bmatrix}a_1^{+} \\ a_1^{-}\end{bmatrix} \\ \overline{\overline{S}}_{\mathrm{PTC}} &= \begin{bmatrix}1/2 & -\eta_{N+1}/2 \\ 1/2 & \eta_{N+1}/2\end{bmatrix}\overline{\overline{M}}_{\mathrm{PTC}}\begin{bmatrix}1 & 1 \\ -1/\eta_1 & 1/\eta_1\end{bmatrix}\end{aligned} \tag{S18}$$

Using the fact that $a_1^{-} = 0$ for incident light, then one has

$$\begin{aligned}\tilde{t}_{\mathrm{PTC}} = \overline{\overline{S}}_{\mathrm{PTC},11} &= \frac{1}{2}\left(\overline{\overline{M}}_{\mathrm{PTC},11} - \overline{\overline{M}}_{\mathrm{PTC},12}/\eta_1\right) - \frac{\eta_{N+1}}{2}\left(\overline{\overline{M}}_{\mathrm{PTC},21} - \overline{\overline{M}}_{\mathrm{PTC},22}/\eta_1\right) \\ \tilde{r}_{\mathrm{PTC}} = \overline{\overline{S}}_{\mathrm{PTC},21} &= \frac{1}{2}\left(\overline{\overline{M}}_{\mathrm{PTC},11} - \overline{\overline{M}}_{\mathrm{PTC},12}/\eta_1\right) + \frac{\eta_{N+1}}{2}\left(\overline{\overline{M}}_{\mathrm{PTC},21} - \overline{\overline{M}}_{\mathrm{PTC},22}/\eta_1\right)\end{aligned} \tag{S19}$$

By following the same procedure, one has the transmission and reflection coefficients (i.e. $\tilde{t}_{\mathrm{MG}}$ and $\tilde{r}_{\mathrm{MG}}$) for the homogenized temporal slab, namely

$$\begin{aligned}\tilde{t}_{\mathrm{MG}} &= \frac{1}{2}\left(\overline{\overline{M}}_{\mathrm{MG},11} - \overline{\overline{M}}_{\mathrm{MG},12}/\eta_1\right) - \frac{\eta_{N+1}}{2}\left(\overline{\overline{M}}_{\mathrm{MG},21} - \overline{\overline{M}}_{\mathrm{MG},22}/\eta_1\right) \\ \tilde{r}_{\mathrm{MG}} &= \frac{1}{2}\left(\overline{\overline{M}}_{\mathrm{MG},11} - \overline{\overline{M}}_{\mathrm{MG},12}/\eta_1\right) + \frac{\eta_{N+1}}{2}\left(\overline{\overline{M}}_{\mathrm{MG},21} - \overline{\overline{M}}_{\mathrm{MG},22}/\eta_1\right)\end{aligned} \tag{S20}$$

On this basis, we obtain the energy transmittivity $\widetilde{T}$ and $\widetilde{R}$ reflectivity. By using the complex Poynting's theorem, the complex Poynting's vector for the incident wave is given by

$$\overline{S}_i = \frac{1}{2}Re\left[\overline{E}_1(t) \times \overline{H}_1^{*}(t)\right] = \hat{k}\frac{\left|a_1^{+}\right|^2}{2\mu_1\sqrt{\varepsilon_1\mu_1}} \tag{S21}$$

where $\hat{k}$ is the unit vector in the direction of the wavevector $\overline{k}$. Similarly, one has the complex Poynting's vector for the transmitted and reflected wave as follows

$$\overline{S}_t = \hat{k}\frac{\left|a_{N+1}^{-}\right|^2}{2\mu_{N+1}\sqrt{\varepsilon_{N+1}\mu_{N+1}}}$$
$$\overline{S}_r = \hat{k}\frac{\left|a_{N+1}^{-}\right|^2}{2\mu_{N+1}\sqrt{\varepsilon_{N+1}\mu_{N+1}}} \tag{S22}$$

Therefore, the energy transmittivity $\widetilde{T}$ and reflectivity $\widetilde{R}$ are related to the transmission and reflection coefficients ($\widetilde{t}$ and $\widetilde{r}$) by

$$\widetilde{T} = \frac{\mu_1\sqrt{\varepsilon_1\mu_1}}{\mu_{N+1}\sqrt{\varepsilon_{N+1}\mu_{N+1}}}\left|\widetilde{t}\right|^2$$
$$\widetilde{R} = \frac{\mu_1\sqrt{\varepsilon_1\mu_1}}{\mu_{N+1}\sqrt{\varepsilon_{N+1}\mu_{N+1}}}\left|\widetilde{r}\right|^2 \tag{S23}$$

## S1.5 Equivalence of the characteristic matrixes between the photonic time crystal and the effective temporal slab

Finally in this subsection, we prove the validity of various Maxwell-Garnett theory, by showing the equivalence of the transmission and reflection coefficients, between the photonic time crystals and their homogenized counterparts, namely

$$\widetilde{t}_{\text{PTC}} = \widetilde{t}_{\text{MG}} \text{ and } \widetilde{r}_{\text{PTC}} = \widetilde{r}_{\text{MG}} \tag{S24}$$

In light of equations (S19) and (S20), it is sufficient to prove equation (S24), if we can obtain

$$\overline{\overline{M}}_{\text{PTC}} = \overline{\overline{M}}_{\text{MG}} \tag{S25}$$

where $\overline{\overline{M}}_{\text{PTC}}$ and $\overline{\overline{M}}_{\text{MG}}$ are the characteristic matrixes for the photonic time crystal and the effective temporal slab, as respectively determined in equation (S14) and (S15). To satisfy equation (S25), one can reasonably expect a stricter condition in the periodic system, namely, the equivalence between the characteristic matrix $\overline{\overline{M}}_{\text{PTC,unit}}$ for each unit cell of the photonic time crystal and that ($\overline{\overline{M}}_{\text{MG,unit}}$) for the homogenized temporal slab of the same thickness, as follows

$$\overline{\overline{M}}_{\text{PTC,unit}} = \overline{\overline{M}}_{\text{MG,unit}} \tag{S26}$$

$$\begin{aligned}
\overline{\overline{M}}_{\text{PTC,unit,11}} &= \cos(\omega_{\text{I}}\tau_{\text{I}})\cos(\omega_{\text{II}}\tau_{\text{II}}) - \sin(\omega_{\text{I}}\tau_{\text{I}})\sin(\omega_{\text{II}}\tau_{\text{II}})\,\eta_{\text{I}}/\eta_{\text{II}} \\
\overline{\overline{M}}_{\text{PTC,unit,12}} &= i\eta_{\text{II}}\cos(\omega_{\text{I}}\tau_{\text{I}})\sin(\omega_{\text{II}}\tau_{\text{II}}) + i\eta_{\text{I}}\sin(\omega_{\text{I}}\tau_{\text{I}})\cos(\omega_{\text{II}}\tau_{\text{II}}) \\
\overline{\overline{M}}_{\text{PTC,unit,21}} &= i/\eta_{\text{I}}\sin(\omega_{\text{I}}\tau_{\text{I}})\cos(\omega_{\text{II}}\tau_{\text{II}}) + i/\eta_{\text{II}}\cos(\omega_{\text{I}}\tau_{\text{I}})\sin(\omega_{\text{II}}\tau_{\text{II}}) \\
\overline{\overline{M}}_{\text{PTC,unit,22}} &= \cos(\omega_{\text{I}}\tau_{\text{I}})\cos(\omega_{\text{II}}\tau_{\text{II}}) - \sin(\omega_{\text{I}}\tau_{\text{I}})\sin(\omega_{\text{II}}\tau_{\text{II}})\,\eta_{\text{II}}/\eta_{\text{I}}
\end{aligned} \tag{S27}$$

$$\overline{\overline{M}}_{\text{MG,unit}} = \begin{bmatrix} \cos\big(\omega_{\text{MG}}(\tau_{\text{I}} + \tau_{\text{II}})\big) & i\eta_{\text{MG}} \sin\big(\omega_{\text{MG}}(\tau_{\text{I}} + \tau_{\text{II}})\big) \\ i \sin\big(\omega_{\text{MG}}(\tau_{\text{I}} + \tau_{\text{II}})\big)/\eta_{\text{MG}} & \cos\big(\omega_{\text{MG}}(\tau_{\text{I}} + \tau_{\text{II}})\big) \end{bmatrix} \tag{S28}$$

Below, we show how equation (S26) is fulfilled under the condition of various Maxwell-Garnett theories.

*For conventional type 1 of Maxwell-Garnett theory within the long-wavelength limit* [3], as derived in equation (4) in the main text, namely

$$\begin{matrix} \dfrac{\tau_{\text{I}} + \tau_{\text{II}}}{\varepsilon_{\text{MG}}} = \dfrac{\tau_{\text{I}}}{\varepsilon_{\text{I}}} + \dfrac{\tau_{\text{II}}}{\varepsilon_{\text{II}}} \\ \dfrac{\tau_{\text{I}} + \tau_{\text{II}}}{\mu_{\text{MG}}} = \dfrac{\tau_{\text{I}}}{\mu_{\text{I}}} + \dfrac{\tau_{\text{II}}}{\mu_{\text{II}}} \end{matrix}, \text{ if } \textit{within} \text{ the long-wavelength limit (including } \omega_{\text{MG}}(\tau_{\text{I}} + \tau_{\text{II}}) \to 0) \tag{S29}$$

By using equation (S29), one can simplify $\eta_{\text{MG}}$ and $\omega_{\text{MG}}$ as

$$\begin{gathered} \eta_{\text{MG}} = \sqrt{\frac{\mu_{\text{MG}}}{\varepsilon_{\text{MG}}}} = \sqrt{\frac{\dfrac{\tau_{\text{I}}}{\varepsilon_{\text{I}}} + \dfrac{\tau_{\text{II}}}{\varepsilon_{\text{II}}}}{\dfrac{\tau_{\text{I}}}{\mu_{\text{I}}} + \dfrac{\tau_{\text{II}}}{\mu_{\text{II}}}}} \\ \omega_{\text{MG}} = \frac{k}{\sqrt{\mu_{\text{MG}}\varepsilon_{\text{MG}}}} = \frac{k}{\tau_{\text{I}} + \tau_{\text{II}}} \sqrt{\left(\frac{\tau_{\text{I}}}{\varepsilon_{\text{I}}} + \frac{\tau_{\text{II}}}{\varepsilon_{\text{II}}}\right)\left(\frac{\tau_{\text{I}}}{\mu_{\text{I}}} + \frac{\tau_{\text{II}}}{\mu_{\text{II}}}\right)} \end{gathered} \tag{S30}$$

Moreover, within the long-wavelength limit, the characteristic matrixes are simplified to

$$\overline{\overline{M}}_{\text{PTC,unit}} = \begin{bmatrix} 1 & i\eta_{\text{II}}\omega_{\text{II}}\tau_{\text{II}} + i\eta_{\text{I}}\omega_{\text{I}}\tau_{\text{I}} \\ i\omega_{\text{I}}\tau_{\text{I}}/\eta_{\text{I}} + i\omega_{\text{II}}\tau_{\text{II}}/\eta_{\text{II}} & 1 \end{bmatrix} \tag{S31}$$

and

$$\overline{\overline{M}}_{\text{MG,unit}} = \begin{bmatrix} 1 & i\eta_{\text{MG}}\omega_{\text{MG}}(\tau_{\text{I}} + \tau_{\text{II}}) \\ i\omega_{\text{MG}}(\tau_{\text{I}} + \tau_{\text{II}})/\eta_{\text{MG}} & 1 \end{bmatrix} \tag{S32}$$

where the Taylor expansion of sine and cosine functions are used. Then, to prove $\overline{\overline{M}}_{\text{PTC,unit}} = \overline{\overline{M}}_{\text{MG,unit}}$ (or more accurately speaking $\overline{\overline{M}}_{\text{PTC,unit}} \approx \overline{\overline{M}}_{\text{MG,unit}}$ in this case), reduces to proving

$$\begin{cases} \eta_{\text{MG}}\omega_{\text{MG}}(\tau_{\text{I}} + \tau_{\text{II}}) = \eta_{\text{II}}\omega_{\text{II}}\tau_{\text{II}} + \eta_{\text{I}}\omega_{\text{I}}\tau_{\text{I}} \\ \omega_{\text{MG}}(\tau_{\text{I}} + \tau_{\text{II}})/\eta_{\text{MG}} = \omega_{\text{I}}\tau_{\text{I}}/\eta_{\text{I}} + \omega_{\text{II}}\tau_{\text{II}}/\eta_{\text{II}} \end{cases} \tag{S33}$$

At this point, equation (S33) can be easily derived through simple algebra based on equation (S30). The detailed mathematics are omitted here.

*For anomalous type 2 of Maxwell-Garnett theory via impedance matching* [4], as governed by equation (6) in the main text, namely

$$\begin{matrix} \dfrac{T_{\text{PTC}}}{\varepsilon_{\text{MG}}} = \dfrac{\tau_{\text{I}}}{\varepsilon_{\text{I}}} + \dfrac{\tau_{\text{II}}}{\varepsilon_{\text{II}}} \\ \dfrac{T_{\text{PTC}}}{\mu_{\text{MG}}} = \dfrac{\tau_{\text{I}}}{\mu_{\text{I}}} + \dfrac{\tau_{\text{II}}}{\mu_{\text{II}}} \end{matrix}, \text{ if } \eta_{\text{I}} = \eta_{\text{II}}, \text{ for } \forall\ \omega_{\text{MG}}T_{\text{PTC}}/2\pi = T_{\text{PTC}}/T_{\text{MG}} \tag{S34}$$

Based on equation (S34), one has

$$\eta_{\mathrm{MG}} = \eta_{\mathrm{I}} = \eta_{\mathrm{II}}$$
$$\omega_{\mathrm{MG}} = \frac{k}{\tau_{\mathrm{I}} + \tau_{\mathrm{II}}}\left(\frac{\tau_{\mathrm{I}}}{\sqrt{\mu_{\mathrm{I}}\varepsilon_{\mathrm{I}}}} + \frac{\tau_{\mathrm{II}}}{\sqrt{\mu_{\mathrm{II}}\varepsilon_{\mathrm{II}}}}\right) = \frac{\omega_{\mathrm{I}}\tau_{\mathrm{I}} + \omega_{\mathrm{II}}\tau_{\mathrm{II}}}{\tau_{\mathrm{I}} + \tau_{\mathrm{II}}} \tag{S35}$$

Moreover, based on the impedance matching condition, namely $\eta_{\mathrm{I}} = \eta_{\mathrm{II}}$, equations (S27) and (S28) respectively reduce to

$$\overline{\overline{M}}_{\mathrm{PTC,unit}} = \begin{bmatrix} \cos(\omega_{\mathrm{I}}\tau_{\mathrm{I}} + \omega_{\mathrm{II}}\tau_{\mathrm{II}}) & i\eta_{\mathrm{I}}\sin(\omega_{\mathrm{I}}\tau_{\mathrm{I}} + \omega_{\mathrm{II}}\tau_{\mathrm{II}}) \\ i\sin(\omega_{\mathrm{I}}\tau_{\mathrm{I}} + \omega_{\mathrm{II}}\tau_{\mathrm{II}})/\eta_{\mathrm{II}} & \cos(\omega_{\mathrm{I}}\tau_{\mathrm{I}} + \omega_{\mathrm{II}}\tau_{\mathrm{II}}) \end{bmatrix} \tag{S36}$$

and

$$\overline{\overline{M}}_{\mathrm{MG,unit}} = \begin{bmatrix} \cos\big(\omega_{\mathrm{MG}}(\tau_{\mathrm{I}} + \tau_{\mathrm{II}})\big) & i\eta_{\mathrm{MG}}\sin\big(\omega_{\mathrm{MG}}(\tau_{\mathrm{I}} + \tau_{\mathrm{II}})\big) \\ i\sin\big(\omega_{\mathrm{MG}}(\tau_{\mathrm{I}} + \tau_{\mathrm{II}})\big)/\eta_{\mathrm{MG}} & \cos\big(\omega_{\mathrm{MG}}(\tau_{\mathrm{I}} + \tau_{\mathrm{II}})\big) \end{bmatrix} \tag{S37}$$

At this point, the equivalence between equations (S36) and (S37) is clear by substituting equation (S35) into them.

*For anomalous type 3 of Maxwell-Garnett theory via temporal Fabry-Pérot*, as governed by equation (8) in the main text, namely

$$\begin{aligned} \frac{T_{\mathrm{PTC}}}{\varepsilon_{\mathrm{MG}}} &= \frac{\tau_{\mathrm{I}}}{\varepsilon_{\mathrm{I}}\eta_{\mathrm{I}}/\eta_{\mathrm{II}}} + \frac{\tau_{\mathrm{II}}}{\varepsilon_{\mathrm{II}}} \\ \frac{T_{\mathrm{PTC}}}{\mu_{\mathrm{MG}}} &= \frac{\tau_{\mathrm{I}}}{\mu_{\mathrm{I}}\eta_{\mathrm{II}}/\eta_{\mathrm{I}}} + \frac{\tau_{\mathrm{II}}}{\mu_{\mathrm{II}}} \end{aligned}, \text{ if } \sin(\omega_{\mathrm{I}}\tau_{\mathrm{I}}) = 0, \text{ for } \omega_{\mathrm{MG}}T_{\mathrm{PTC}}/2\pi = T_{\mathrm{PTC}}/T_{\mathrm{MG}} > 1/2 \tag{S38}$$

Based on equation (S38), one has

$$\eta_{\mathrm{MG}} = \eta_{\mathrm{II}}$$
$$\omega_{\mathrm{MG}} = \frac{k}{\tau_{\mathrm{I}} + \tau_{\mathrm{II}}}\left(\frac{\tau_{\mathrm{I}}}{\sqrt{\mu_{\mathrm{I}}\varepsilon_{\mathrm{I}}}} + \frac{\tau_{\mathrm{II}}}{\sqrt{\mu_{\mathrm{II}}\varepsilon_{\mathrm{II}}}}\right) = \frac{\omega_{\mathrm{I}}\tau_{\mathrm{I}} + \omega_{\mathrm{II}}\tau_{\mathrm{II}}}{\tau_{\mathrm{I}} + \tau_{\mathrm{II}}} \tag{S39}$$

Furthermore, based on the temporal Fabry-Pérot resonance condition, e.g. $\sin(\omega_{\mathrm{I}}\tau_{\mathrm{I}}) = 0$, equation (S27) reduces to

$$\begin{aligned} \overline{\overline{M}}_{\mathrm{PTC,unit}} &= \begin{bmatrix} (-1)^m\cos(\omega_{\mathrm{II}}\tau_{\mathrm{II}}) & i\eta_{\mathrm{II}}(-1)^m\sin(\omega_{\mathrm{II}}\tau_{\mathrm{II}}) \\ i/\eta_{\mathrm{II}}\,(-1)^m\sin(\omega_{\mathrm{II}}\tau_{\mathrm{II}}) & (-1)^m\cos(\omega_{\mathrm{II}}\tau_{\mathrm{II}}) \end{bmatrix} \\ &= \begin{bmatrix} \cos(m\pi + \omega_{\mathrm{II}}\tau_{\mathrm{II}}) & i\eta_{\mathrm{II}}\sin(m\pi + \omega_{\mathrm{II}}\tau_{\mathrm{II}}) \\ i\sin(m\pi + \omega_{\mathrm{II}}\tau_{\mathrm{II}})/\eta_{\mathrm{II}} & \cos(m\pi + \omega_{\mathrm{II}}\tau_{\mathrm{II}}) \end{bmatrix} \end{aligned} \tag{S40}$$

where the identities of $\cos(\omega_{\mathrm{I}}\tau_{\mathrm{I}}) = (-1)^m$ and $(-1)^m\cos(\omega_{\mathrm{II}}\tau_{\mathrm{II}}) = \cos(m\pi + \omega_{\mathrm{II}}\tau_{\mathrm{II}})$ are used. By use the Fabry- Pérot resonance condition again, namely $\omega_{\mathrm{I}}\tau_{\mathrm{I}} = m\pi$, one has

$$\overline{\overline{M}}_{\mathrm{PTC,unit}} = \begin{bmatrix} \cos(\omega_{\mathrm{I}}\tau_{\mathrm{I}} + \omega_{\mathrm{II}}\tau_{\mathrm{II}}) & i\eta_{\mathrm{II}}\sin(\omega_{\mathrm{I}}\tau_{\mathrm{I}} + \omega_{\mathrm{II}}\tau_{\mathrm{II}}) \\ i\sin(\omega_{\mathrm{I}}\tau_{\mathrm{I}} + \omega_{\mathrm{II}}\tau_{\mathrm{II}})/\eta_{\mathrm{II}} & \cos(\omega_{\mathrm{I}}\tau_{\mathrm{I}} + \omega_{\mathrm{II}}\tau_{\mathrm{II}}) \end{bmatrix} \tag{S41}$$

Similarly, at this point, the equivalence between the characteristic matrix for the unit cell of the photonic time crystal in equation (S40), and that of the homogenized temporal slab of the same thickness can be easily obtained, based on equation (S39).

From all above, we have prove the equivalence between the transmission coefficient $\tilde{t}_{\text{PTC}}$ (or the reflection coefficient $\tilde{r}_{\text{PTC}}$) for a temporally finitely-thick photonic time crystal and that ($\tilde{t}_{\text{MG}}$ or $\tilde{r}_{\text{MG}}$) for the effective temporal slab, namely $\tilde{t}_{\text{PTC}} = \tilde{t}_{\text{MG}}$ (or $\tilde{r}_{\text{PTC}} = \tilde{r}_{\text{MG}}$), in a strict manner, by showing the equivalence between their character matrixes.

## S2 Spatiotemporal evolution of various wave packets interacting with photonic time crystals beyond the long-wavelength limit.

In this section we give the rigorous expressions for the field distribution of various space-time wave packet interacting with the photonic time crystal beyond the long-wavelength limit. The incident wave packet takes the form

$$\begin{bmatrix} a_1^+ \\ a_1^- \end{bmatrix} = \begin{bmatrix} a(k) \\ 0 \end{bmatrix} \tag{S42}$$

where $a(k)$ is the wavevector-dependent amplitude of the space-harmonic wave packet. For example, for the continuous Gaussian-type waveform, $a(k) = e^{-k^2/2\sigma_k^2}$. On this basis, one can obtain the field amplitude in region $j$, for $\forall j \in [2, N+1]$.

$$\begin{bmatrix} a_j^+ \\ a_j^- \end{bmatrix} = \begin{bmatrix} 1/2 & -\eta_j/2 \\ 1/2 & \eta_j/2 \end{bmatrix} \cdot \left[ \prod_{n=j-1}^{1} \overline{\overline{M}}_n \right] \cdot \begin{bmatrix} 1 & 1 \\ -1/\eta_1 & 1/\eta_1 \end{bmatrix} \begin{bmatrix} a(k) \\ 0 \end{bmatrix}, \quad \forall j \in [2, N+1] \tag{S43}$$

By substituting the values of $a_j^+$ and $a_j^-$ into equation (S3), all the spatiotemporal evolution of the wave packet can be obtained.